\let\new=\newcommand
\new{\eq}{\begin{equation}}
\new{\en}{\end{equation}}
\new{\degree}{$^\circ$ }
 
\documentclass[12pt,preprint2]{aastex}

\shorttitle{Duplicity Survey of Exoplanet Host Stars}
\shortauthors{Roberts et al.}

\begin{document}


\title{Know the Star, Know the Planet. I. \\ Adaptive Optics of Exoplanet Host Stars}

\author{Lewis C. Roberts, Jr.}  
\affil{Jet Propulsion Laboratory, California Institute of Technology, 4800 Oak Grove Drive, Pasadena CA 91109, USA}
\email{lewis.c.roberts@jpl.nasa.gov}

\author{Nils H. Turner, Theo A. ten Brummelaar}
\affil{Center for High Angular Resolution Astronomy, Georgia State University, Mt. Wilson, CA 91023, USA}
\email{nils@chara-array.org, theo@chara-array.org}

\author{Brian D. Mason, William I. Hartkopf}
\affil{U.S. Naval Observatory, 3450 Massachusetts Avenue, NW, Washington, DC 20392-5420, USA}
\email{bdm@usno.navy.mil, wih@usno.navy.mil}


\begin{abstract}

The results of an adaptive optics survey of exoplanet host stars for stellar
companions is presented. We used the AEOS Telescope and its adaptive optics 
system to collect deep images of the stars in $I$-band. Sixty-two exoplanet 
host stars were observed and fifteen multiple star systems were resolved. Of
these eight are known multiples, while seven are new candidate binaries. For
all binaries, we measured the relative astrometry of the pair and the 
differential magnitude in $I$-band. We improved the orbits of HD 19994 and 
$\tau$ Boo. These observations will provide improved statistics on the 
duplicity of exoplanet hosts stars and provide an increased understanding of
the dynamics of known binary star exoplanet hosts. 

\end{abstract}



\textbf{Accepted to Astronomical Journal}

\section{Introduction}

There are a number of reasons to gain a better understanding of the binary fraction of exoplanet hosts. Binary stars influence the environment of the planet, both affecting where planets form and altering their orbits. Unseen binary stars complicate the analysis of exoplanets, and under some circumstances may masquerade as exoplanets themselves. 

There are several proposed ways in which binary stars can influence the orbital properties of exoplanets. Binaries may interact with the protoplanetary disk from which planets form; for binary stars with separations less than 100 AU, interactions with the protoplanetary disk may lead to altered planet formation. If the protoplanetary disk surrounds the primary star, the tidal torques of the companion may  lead to a truncation of the disk \citep{artymowicz1994}.  Eccentric companions can cause the disk to assume a similar shape, leading to a more eccentric orbit of the planet, which is then driven inwards while accreting large amounts of gas \citep{kley2010}.  Kley \& Nelson also suggest the companion may alter the planetesimal accretion rates in a disk with a close binary companion.  Core accretion may have difficulties explaining planets orbiting one member of a close binary system; an example of this is the $\gamma$ Cep system. 
  
The problem with validating these theories and models is that we are unable to observe the process as it occurs. Instead, we are forced to work backwards, using statistical analysis of duplicity fraction and orbital parameters to determine what happened. There have been several analyses of the statistics of binary stars. According to a study by \citet{eggenberger2004}, massive planets in close orbits are mostly found in binaries. They also noted other possible peculiar characteristics of planets in binaries: low eccentricities for orbital periods shorter than 40 days and a lack of massive planets with periods longer than 100 days in binaries.  Within three years of their analysis, the number of planets in binaries increased from 19 planets in 15 multiple star systems to 40 planets in 38 multiple star systems. \citet{desidera2007} studied the distribution of eccentricity, mass, period and metallicity of planets in this larger sample of binary systems. They found that the \citet{eggenberger2004} suggestion that there was a lack of massive planets with periods $>$100 days in binaries was not valid. Instead, massive planets in short period orbits are found in most cases around the components of close binaries.  \citet{raghavan2006} carried out a comprehensive review of the literature and archival data to show that for 23\% of radial velocity (RV) detected planets, the host star was a binary and that three planetary systems were in triple systems with the possibility that another two were also in triple systems.

Undetected binaries or even optical doubles can alter the analysis of exoplanets. For transiting exoplanets, the transit depth is assumed to be a percentage of a single star's brightness, but an undetected companion or nearby field star can affect the accuracy of the results. Correcting for this can change the derived planetary diameters by a few percent \citep{daemgen2009}.   
 
Exoplanet host stars are also prime targets for coronagraphic searches for additional exoplanets (e.g. Leconte et al. 2010).  Identifying background stars and stellar companions assist those observations by eliminating the need to spend observing time establishing the true nature of each candidate planet.  

Finally, some candidate exoplanets are actually stellar companions. This misidentification usually arises from radial velocity observations, where the mass and semi-major axis results are projected onto the inclination of the orbit. 
A low-mass star in a nearly face-on orbit can masquerade as a massive planet in a higher-inclination orbit.  This will probably be an increasing problem as the time baseline for RV searches increases.  An example of such a masquerading star is HD 104304. \citet{schnupp2010} used adaptive optics (AO) and ``Lucky Imaging'' to show that the substellar object orbiting this solar analog was actually an M4V star in a near face-on orbit. Another example is HD 8673, which is probably a K or M star in a low inclination orbit \citep{mason2011}.

It is essential to eliminate masquerading stars when analyzing the statistics of exoplanets because it prevents the unnecessary application of observing time and other resources towards better understanding an ``exoplanet'', when in actuality it is a star.  It also offers an opportunity for stellar astronomy, in that it is possible to compute the masses of the individual stars by combining spectroscopic and astrometric orbital information.  There are relatively few well determined stellar masses and any opportunity to increase this number is valuable.

There have been a number of prior papers that studied duplicity rates among the exoplanet hosts. \citet{patience2002} used near-IR speckle interferometry and AO to observe 11 exoplanet hosts and discovered one binary and collected astrometry on two known binaries.  Southern exoplanet hosts were observed by Eggenberger et al. (2007) and in the 57 exoplanet hosts and 73 control stars they observed, they found 19 true companions, 2 likely bound objects, and 34 background stars.  Another 40 candidate companions require additional data to determine if they are physical.  \citet{chauvin2006} used the Very Large Telescope (VLT) and Canada France Hawaii Telescope (CFHT) to study 26 stars with planets and found 3 confirmed companions and 8 candidate companions, which also require follow up observations.  Mugrauer and colleagues have observed a number of stars in a large series of papers that cover the detection of a few binaries in each paper (e.g., Mugrauer et al. 2005, 2007, Neuh\"auser et al. 2007). 

With these thoughts, we started a survey of exoplanet hosts in 2001 with the Advanced Electro-Optical System (AEOS) telescope and  its AO system.   Section \ref{data} discusses the observations and data reduction methodology.  Results are discussed in Section \ref{results}. Subsections are dedicated to observations of new candidate companions, null detections and the computation of orbits to previously known  exoplanet host stars. Section \ref{2mass} discusses archival observations from 2MASS and finally Section \ref{summary} summarizes our results. 


\section{Data Collection and Analysis}\label{data}

Observations were made using the  AEOS 3.6m telescope and its AO system.  The AEOS telescope is located at the Maui Space Surveillance System at the summit of Haleakala.  There were dedicated observing runs in 2001 February, 2001 September, 2002 March and 2002 September.  Stars were also observed on a queue scheduled basis between 2001 and 2005.  

The AEOS AO system is a natural guide star system using a Shack-Hartmann wavefront sensor \citep{roberts2002}.   The individual subaperatures have a diameter of 11.9 cm projected onto the primary.   The deformable mirror has 941 actuators.  The system's closed loop bandwidth is adjustable and can run up to 200 Hz, although the normal range is approximately 50 Hz.  In the configuration used for these observations, the light from  500-540 nm is sent to the tip/tilt detector system, the light from 540-700 nm is sent to the  wavefront sensor and the light longer than 700 nm is sent to the Visible Imager CCD science camera. It operates from 700 to 1050 nm; is equipped with an atmospheric dispersion corrector; has a two-mode image derotator (zenith at a fixed position in the image or celestial north at a fixed position in the image); and, for this project, has a 10\arcsec~FOV (0\farcs022 pixel$^{-1}$).   The detector is a 512$\times$512 pixel E2V CCD, with  a read noise of 12 e$^-$ RMS. The camera output is digitized to 12 bits with 10 e$^-$ per digital number.
 
The observing list was created by taking the list of known stars with known exoplanets as of early 2001 and then removing stars fainter than  the effective magnitude limit of the AEOS AO system (about m$_v$ = 8), and stars that were outside of the declination limit of acceptable AO correction (objects that at some time during the year get above 30\degr ~elevation   at Haleakala, i.e., a declination greater than about -45\degr). In addition, we had an ongoing program to observe known binaries with the AEOS telescope \citep{roberts2011}.  We compared the list of observed target against the list of known exoplanet host stars to find targets that were not known to host an exoplanet at the time we observed them.  This turned up several additional objects, which are included in this paper. 

Each data set consists of 1000 frames obtained  using a Bessel $I$-band filter ($\lambda_0$ = 880nm). After collection, any saturated frames are discarded and the remaining frames are debiased, dark subtracted and flat fielded.  The frames are weighted by their peak pixel, which is proportional to their Strehl ratio and then co-added using a shift-and-add routine. The resulting image is analyzed with the program \textit{fitstars}; it uses an iterative blind-deconvolution that fits the location of delta functions and their relative intensity to the data.  The co-adding technique and the analysis with \textit{fitstars} was presented in ten Brummelaar et al. (1996, 2000).

Error bars on the astrometry and photometry were assigned using the method described  in \citet{roberts2005}. For the photometry, simulated binary stars were created from observations of single stars.  The photometry of these simulated binaries was measured and used to create a grid of measurement errors as a function of separation and differential magnitudes.  For astrometry, the separation error bar is $\pm$0\farcs02 for $\rho$ $\leq$ 1\arcsec, $\pm$0\farcs01 for 1\arcsec $<$ $\rho$ $\leq$ 4\arcsec, and $\pm$0\farcs02 for $\rho$ $>$ 4\arcsec.  The position angle error bar is $\pm2^\circ$ for $\rho < 1$\arcsec and $\pm1^\circ$ for $\rho > 1$\arcsec. The Visible Imager Data taken in 2001 did not record which state the derotator was in (either fixed North or fixed Zenith).  As such there is an ambiguity in the position angle for these measurements.

\section{Results}\label{results}

In Section \ref{known_binaries}, we discuss the observations of known binaries, while in Section \ref{new_binaries} we discuss binaries detected for the first time. Finally in Section \ref{single}, we discuss the stars where no companion was detected.  The astrometry and photometry of all resolved systems are listed in Table \ref{exo_binaries}. 

For each star, we list the Washington Double Star (WDS) number, the discovery designation, the most common planet designation if it has one other than the HD number, the Hipparcos and HD Catalogue numbers, the Besselian date of the observation, the separation in arcseconds, the position angle in degrees and finally the differential magnitude measured in Bessel $I$-band. The listed astrometry was compared with the latest published astrometry in the WDS. In some cases this helped alleviate the derotator discrepancy between zenith mode and astronomical north mode and allowed us to determine the position angle.  In all cases, our astrometry was consistent with previous results.  

\subsection{Known Binaries}\label{known_binaries}
 
\textit{HD 19994 (WDS 03128-0112)}  

\begin{figure}[tb]
\includegraphics[width=84mm]{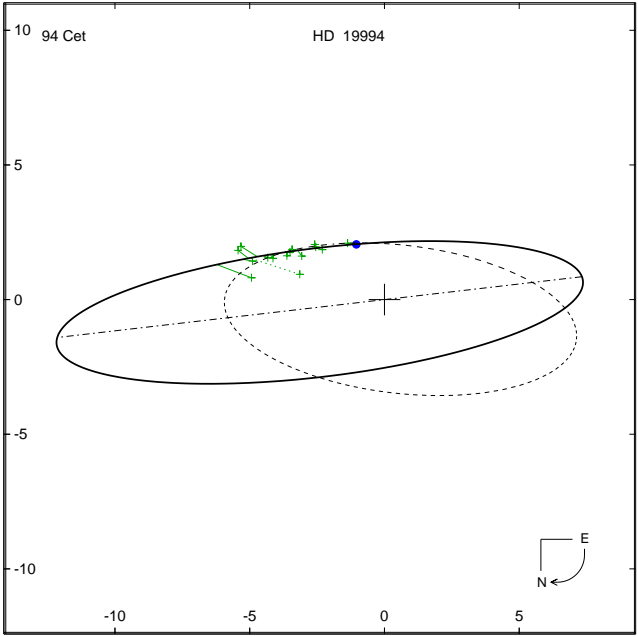}
\caption{New orbit for the binary star HD 19994.  The orbit with dashed lines is that of \citep{hale1994}.  The broken line through the   origin is the line of nodes. The scale of the orbit in arcseconds is given at left and bottom and the orientation and direction of   motion is at lower right. The `+' signs are historic filar   micrometry measures and the filled circle is the new AO measure  listed in Table \ref{exo_binaries}. Measures are connected to the new orbit by O-C lines. 
}
\label{hj663_orbit}
\end{figure}

This system has a 1.68 M$_J$ planet in a 535$\pm$3.10 day period \citep{mayor2004}. Its semi major axis is 1.42 AU. We examined the  previous orbit for this system \citep{hale1994} and determined a new orbit was warranted.  There is a large scatter in measured separation in the early measurements, which results in a poor solution. Convergence was achieved by fixing the eccentricity to a value minimizing the O-C of measures and then allowing other parameters to vary. With all  parameters floating, the orbit solution quickly diverged with   extremely high eccentricity. It is possible that the B companion is not physical and that early measures indicating curvilinear motion are measures with higher error to a linear fit. At present, there  is no differential proper motion of the B   component to confirm or deny this supposition. The orbit  solution, an improvement on the solution of   Hale (1994), should   still be judged as preliminary. The elements of the orbit are shown in Table \ref{orbital_elements}, while a plot of the orbit is shown in Figure \ref{hj663_orbit}.  The predicted orbital positions until 2035 are shown in Table \ref{ephemeris}. The parallax of 44.29 mas \citep{vanLeeuwen2007} implies a mass sum of 2.6 M$_{\sun}$, which is high for the estimated spectral types of the components of F8V and M3V \citep{hale1994}.  The spectral type estimates would suggest a mass sum of 1.6 M$_{\sun}$.  The mass sum from the \citet{hale1994} orbit is 1.8 M$_{\sun}$.

Combining the orbital elements with the parallax, we compute a  separation of the two stars at periastron passage of 163 AU. This is a large enough separation that the companion probably has little or no impact on the planet itself, but if the planet scattered large numbers of planetesimals as it migrated, these planetesimals would then be scattered again by the stellar companion.  This would certainly modify or possibly eliminate an outer debris disk analogous to the Kuiper belt. No debris disk was detected by \citet{dodson-robinson2011}.

\textit{$\tau$ Boo (HD 120136, WDS 13473+1727)} 

\begin{figure}[tb]
\includegraphics[width=84mm]{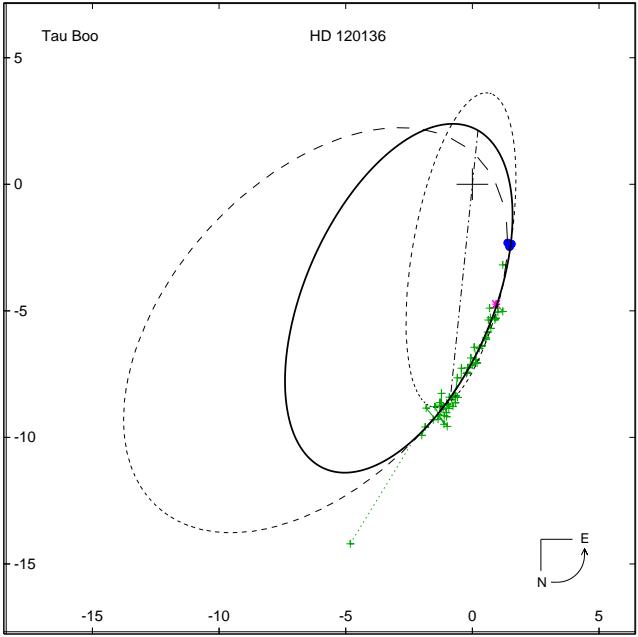}
\caption{New orbit for the stellar companion to $\tau$ Boo.  The orbit with long dash lines is that of \citep{hale1994}, while the orbit with the short dashed lines is the orbit of \citep{popovic1996} The `+' signs are historic filar   micrometry measures and the filled circles are modern speckle interferometry and AO measurements.
}
\label{tauboo_orbit}
\end{figure}

This system hosts a massive planet in a 3.3128 day orbit \citep{butler1997} with a semi-major axis of 0.042 AU. There have been two previous orbits computed for this system, by  \citet{hale1994} and \citet{popovic1996}. We computed a new solution, as shown in Figure \ref{tauboo_orbit}; elements are listed in Table \ref{orbital_elements}. The predicted orbital positions until 2035 are shown in Table \ref{ephemeris}.

The parallax of 64.03 mas \citep{vanLeeuwen2007} implies a mass sum of 2.0 M$_{\sun}$. Hale's orbit gives 2.8  M$_{\sun}$, while that of Popovic \& Pavlovic  gives 6.3  M$_{\sun}$. The estimated spectral types from \citet{hale1994}, F6IV and M2V, results in a mass sum of 1.8 M$_{\sun}$, within 10\% of our estimate.  While our orbit fits the early micrometry data better than the Popovic orbit, and the later adaptive optics and speckle interferometry data better than the Hale orbit, it warrants  further refinement, which requires additional observations.  Although it is moving faster and approaching periastron, given the orbital period, these additional observations will need to be several years in the future.

Combining the orbital elements with the parallax, we compute the separation of the two stars at periastron passage as 30.0 AU. If we assume the binary orbit has not altered through a close stellar encounter, since the formation of the planetary system, the minimum separation has a significant impact on the formation of the planet. This seems like a perfect example of what \citep{kley2010} proposed; eccentric  stellar companions can cause the protoplanetary disk to become more eccentric, which leads to a more eccentric orbit for the planet. The planet is then driven inwards, while accreting large amounts of gas, producing a massive planet. In this case the planet's orbit was probably then tidally circularized. Planetary migration may explain why $\tau$ Boo only has a single massive planet in a very close orbit, as other proto-planets would have been scattered out of the disk. To determine the validity of this idea, detailed numerical simulations will be needed. A more refined orbit will also be useful, which requires additional astrometric measurements.

\subsection{New Discoveries}\label{new_binaries}

\textit{HD 37124} - We detcted two candidate companions with separations of 3\farcs03 and 3\farcs17.  They probably do not form a hiearchal systems as the separations are too similar and one or both of the companions is a foreground or background object. Based on the parallax and assuming face-on orbits, the companions would have separations of 102 AU and 106 AU. The large

\textit{HD 59686} - We detected a candidate companion with a separation 5\farcs61 and a $\Delta$I of 4.6.  Based on the measured separation and the parallax of the star \citep{vanLeeuwen2007} and assuming a face-on orbit with an inclination of zero, the candidate companion would have a minimum separation of 519 AU.

\textit{HD 89744} - A faint candidate companion was detected with a separation of 5\farcs62 and a $\Delta$I of 13$\pm$2.   Based on the spectral type of the primary \citep{montes2001}, the differential magnitude of the companion would make it a brown dwarf.  The star already has one low-mass stellar or  high-mass  brown dwarf companion detected  \citep{mugrauer2004}.  That paper also detected several background or foreground objects that did not share common proper motion with HD 89744. Both our and  \citet{mugrauer2004} observations were made in 2002, so it is doubtful that the object we have detected is one of the  background or foreground objects as they do appear to have similar astrometry.   \citet{mugrauer2004} did not detect our candidate companion, but their observations were not made with AO and probably did not have the dynamic range to detect it.  Follow up  observations   with near-IR AO are needed to make a final determination if our object is a true companion, an artifact or a background object. Again, assuming a face-on orbit, based on the measured separation and the parallax of the star \citep{vanLeeuwen2007}, the candidate companion would have a minimum separation of 219 AU.


\textit{70 Vir (HD 117176)} - \citet{patience2002} did not find a companion to this star in their infrared survey.  That may indicate that this is a false detection, that it is background blue star, or that the orbital position has changed.   Based on the spectral type of the primary \citep{cowley1967}, the absolute magnitudes of the MK classification in \citet{cox2000}, and the measured differential magnitude of 11.4$\pm$1.2 the companion would be later than M5V, assuming it is on the main sequence.  Due to its relatively high proper motion \citep{vanLeeuwen2007}, an additional observation at a current epoch will be able to determine  whether the companion shares   common proper motion or  is  a background object.  Using the same assumptions as above, the candidate companion would have a minimum separation of 52 AU.


\textit{14 Her (HD 145675) } - Based on the spectral type of the primary \citep{montes2001}, the absolute magnitudes of the MK classification in \citet{cox2000},   and  the measured differential magnitude of 10.9$\pm$1.0,  the companion would be later  than  M5V,  assuming it is on the main sequence. Using the same assumptions as above, the candidate companion would have a minimum separation of 78 AU.


\textit{HD 187123 } - Four possible candidates were detected in two images. The first image showed 4 candidate companions, while the second observation  a month  later showed only the brightest  of these possible companions.  The second observation had poorer AO correction which offers a probable explanation for the failure to detect the other candidates.  Based on the spectral type of the primary \citep{gray2001}, the absolute magnitudes of the MK classification in \citet{cox2000}, and the measured differential magnitudes, all three companions would be later  than M5V,  assuming they are on the main sequence. Based on the measured separations and the parallax of the star \citep{vanLeeuwen2007}, the candidate companions would have separations of 127 AU, 143 AU, 155 AU, and 276 AU.  The system does not appear to form a hierarchical system, and it is highly likely that some or all of the companions are background objects.


\textit{51 Peg (HD 217014)} - Using the same assumptions as above, the candidate companion would have a separation of 45 AU. Due to its  relatively high proper motion {vanLeeuwen2007}, an additional observation at a current epoch will be able to determine if the star has common proper motion or is a background object.


\subsection{Single Stars}\label{single}
 
Table  \ref{single_exostars} lists the stars where no companions were detected.  The tables lists the WDS number of the system if there is one,  HD and Hipparcos catalogue numbers,  the observation date and the full width half maximum (FWHM) of the star.  This provides a useful gauge for estimating the minimum separation resolvable by the observations and also provides  a metric  of AO performance. The AO performance varies from night to night, as a function of atmospheric conditions, target brightness and air mass.  

In order to quantify the sensitivity of the reduced images, we have created a variation of the “dynamic range map” technique described in \citet{hinkley2007}, which defines the dynamic range of a given position in a 2D image as the faintest companion detectable at that position to the 5$\sigma$ level. In our version, we construct a map the same size as each reduced image. The intensity level of each pixel in the map is set to five times the root mean square (RMS) intensity variation across a patch centered on the corresponding pixel in the original image. The patch is a square with lengths equal to the FWHM of the original image. This produces the dynamic range map in intensity terms, which are then converted to magnitudes.

\begin{figure}[tb]
\includegraphics[width=84mm]{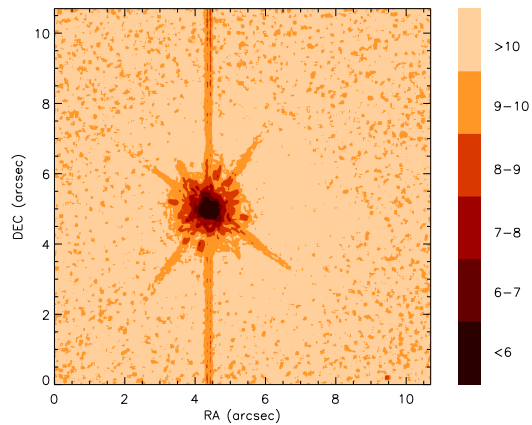}
\caption{ A sample dynamic range map of HD 8574. The gradient map on the right shows the 5σ detection limit for a given pixel of the image in magnitudes.
}
\label{dynamic_range}
\end{figure}

Figure \ref{dynamic_range} shows the results of this technique applied to the image of HD 8574 and is characteristic of the achievable dynamic range. It is apparent from the sample figure that the detection threshold for a companion is highly spatially variant. The dynamic range increases with increasing radius from the central star. There are several artifacts in this image which lower the dynamic range. The largest is the vertical line, which is an artifact of the  shutter-less frame-transfer  process. There is also the diffraction pattern caused by the secondary mirror support spiders.  

 As the AO system performance decreases,  the FWHM  of the point spread function (PSF) will increase, which increases the area over which the central PSF causes confusion. In addition, it also widens the companion’s PSF, lowering the contrast between the two and decreasing the detection rate.  

These maps have several purposes. For those objects with multiple observations where companions are seen in some but not all of the data, changes in the dynamic range can help explain the reasons for these discrepancies and can put constraints on the orbital solution \citep{hinkley2011}.  They also illustrate the limitations of the data, showing where it is almost impossible to find faint companions.  

\subsection{Unresolved Binary Stars}
 
In this section, we discuss two binary stars that based upon the published astrometry, we should have been able to resolve, but did not.  There are other binary stars listed in Table \ref{single_exostars}, but their published separations are either outside of the field of view, or are smaller than the PSF FWHM of the observation.  The minimum separation in Table \ref{single_exostars}  and the widest routinely measureable separation of 5\arcsec gives a  useful `face-on donut' of non-detection.

\textit{HD 217107 (WDS 22583-0224, CHR 116)} - CHR 116 has only been resolved twice: in 1982 and in 1997.  In the 15 years, the separation changed from 0\farcs473 to 0\farcs274. It is quite likely that in the 4 years  between the last measurement and our measurement, orbital motion has moved the companion to within the 0\farcs19 measured FWHM of the image.   Examining the image, we do see a slight elongation of the PSF in one direction. It is possible that this is caused by the secondary.  With a FWHM of 0\farcs19, the AO correction  was not very good. 

\textit{$\gamma$ Cep (HD 222404, WDS 23393+7738, NHR 9)} - We did not detect the stellar companion  seen by \citet{neuhauser2007}. According to the orbit of \citet{torres2007}, the companion would have a separation of 0\farcs17.  As shown in Fig. \ref{dynamic_range}, it would be extremely difficult to detect a companion with the expected dynamic range at that separation.

\section{2MASS Analysis}\label{2mass}

A check was made against the 2MASS Point Source Catalog \citep{skrutskie2006}, looking  for all objects within 10\arcsec~of the target stars as discussed in \citet{turner2008}.    Only three such possible companions  were found, all of which are previously known. The candidate companions to HD 114729 and HD 168746 are outside of our field of view 

79 Cet : A companion was found at 187$^{\circ}$\llap.5 and 5\farcs70, with $\Delta$J = 3.59. \citet{mugrauer2005} detected this companion with AO in 2002 and noted the match with the 2MASS object. 

HD 114729 : This object is listed as a single star in Table \ref{single_exostars}. A possible companion was found in the 2MASS database, but with a separation of about 8\arcsec,  this object falls outside the observing window of the VisIm detector, and we did not detect it.

HD 168746 : This star is also listed as single in Table \ref{single_exostars}. A possible companion was found in the 2MASS database with a separation of 9\arcsec, it is well outside our observing window and was not detected.

\section{Summary}\label{summary}

We observed 62 exoplanet host stars with the AEOS AO system, and resolved 15 multiple star systems. Of these  eight are known multiples, while seven are new candidate binaries.  Additional observations are needed to determine if these are true binaries or merely optical doubles.  We computed updated orbits of HD 19994 and $\tau$ Boo. Both are improved from the previous orbit, but require additional observations to further  refine these solutions. Both orbits show that the presence of a binary companion can affect the structure of the planetary system. In the case of HD 19994, the companion may have disrupted any Kuiper belt analogs. The $\tau$ Boo stellar companion may have altered the protoplanetary disk and caused only a single massive planet to form at a very small separation. In this way, determining the orbits of binary star companions to exoplanet hosts may shed new light on the formation of the exoplanets.

\acknowledgments

We thank the numerous staff members of the Maui Space Surveillance System who helped make  these observations  possible. The US Air Force provided the telescope time, on-site support and 80\% of the research funds for this AFOSR and NSF jointly sponsored research under grant number NSF AST 0088498. L.C.R. was funded by AFRL/DE (Contract Number F29601-00-D-0204) and by the Jet Propulsion Laboratory, California Institute of Technology, under a contract with the National Aeronautics and Space Administration. T.A.tB was supported by the Center for High Angular Resolution Astronomy at Georgia State University.  This research made use of the Washington Double Star Catalog,  maintained at the U.S. Naval Observatory, the SIMBAD database, operated by the CDS in Strasbourg, France and NASA's Astrophysics Data System. This publication makes use of data products from the Two Micron All Sky Survey, which is a joint project of the University of Massachusetts and the Infrared Processing and Analysis Center/California Institute of Technology, funded by the National Aeronautics and Space Administration and the National Science Foundation



\clearpage

\begin{deluxetable}{clcrrcccc}
\tabletypesize{\footnotesize}
\tablewidth{0pt}
\tablecaption{Exoplanet Host Binaries\label{exo_binaries}}
\tablehead{\colhead{WDS} & \colhead{Discovery} & \colhead{Planet} & \colhead{HIP} & \colhead{HD} & \colhead{Epoch} & \colhead{$\rho$}& \colhead{$\theta$}  & \colhead{$\Delta I$} \\ 
\colhead{\#} & \colhead{Designation} & \colhead{ } & \colhead{\#} & \colhead{\#} 
& \colhead{} & \colhead{(\arcsec)}&  \colhead{(\degr)} & }
\startdata
02104$-$5049 & ESG 1                     & GJ 86       &  10138 &  13445 & 2001.7454 & 1.70 & 114.8Z / 112.1N &  $\:$9.0$\pm$0.8 \\

02353$-$0334 & MUG 2 AB                  & 79 Cet      &  12048 &  16141 & 2001.7454 & 6.07 & 191.8Z / 183.7N &  5.64$\pm$0.07 \\

03128$-$0112 & HJ 663                    & \nodata     &  14954 &  19994 & 2002.6845 & 2.32 & 206.8           &  4.33$\pm$0.02 \\ 

05370$+$2044 & RBR 15 AB                 & \nodata     &  26381 &  37124 & 2003.9108 & 3.03 & 160.3           &  9.6$\pm$0.7 \\ 
05370$+$2044 & RBR 15 AC                 & \nodata     &  26381 &  37124 & 2003.9108 & 3.17 & 299.1           &  9.5$\pm$0.7 \\ 
07318$+$1705 & RBR 16                    & \nodata     &  36616 &   59686& 2004.0477 & 5.61 & 224.8           &  4.60$\pm$0.01 \\
10222$+$4114 & RBR 17 AH                 & \nodata     &  50786 &  89744 & 2002.0166 & 5.62 &  26.6Z /  53.6N & 13$\pm$2$\:\:$ \\ 
13123$+$1731 & PAT  47                   & \nodata     &  64426 & 114762 & 2002.1069 & 3.27 &  26.9           &  9.2$\pm$0.7 \\
13284$+$1347 & RBR 18 AD                 & 70 Vir      &  65721 & 117176 & 2001.0994 & 2.86 & 191.3Z / 241.2N & 11.4$\pm$1.2$\:\:$ \\

13473$+$1727 & STT 270                   & $\tau$ Boo  &  67275 & 120136 & 2001.0994 & 2.71 &  31.3           &  5.01$\pm$0.04 \\

16104$+$4349 & RBR 19                    & 14 Her      &  79248 & 145675 & 2002.2382 & 4.32 & 209.0           & 10.9$\pm$1.0$\:\:$ \\

19053$+$2555 & EGN 24                    & \nodata     &  93746 & 177830 & 2002.5474 & 1.62 & 84.1            &  7.5$\pm$0.3 \\

19418$+$5032 & TRN 4Aa,Ab                & 16 Cyg      &  96895 & 186408 & 2002.5502 & 3.38 & 204.7           &  7.6$\pm$0.3 \\

             &                           &             &        &        & 2002.6842 & 3.41 & 204.4           &  7.1$\pm$0.3 \\

19470$+$3425 & RBR 20 AB & \nodata     &  97336 & 187123 & 2002.5503 & 2.65 & 343.6           &  9.3$\pm$0.7 \\

19470$+$3425 & RBR 20 AC & \nodata     &  97336 & 187123 & 2002.5503 & 2.99 & 339.0           &  9.9$\pm$0.7 \\

19470$+$3425 & RBR 20 AD & \nodata     &  97336 & 187123 & 2002.5503 & 3.24 &  67.0           &  7.8$\pm$0.3 \\

             &                           &             &        &        & 2002.6842 & 3.25 &  65.7           &  7.6$\pm$0.3 \\ 

19470$+$3425 & RBR 20 AE & \nodata     &  97336 & 187123 & 2002.5503 & 5.77 & 187.7           & 10.5$\pm$1.0$\:$ \\

20283$+$1846 & HO  131AB                 & \nodata     & 100970 & 195019 & 2002.4902 & 3.53 & 331.3           &  $\:$3.23$\pm$0.01 \\

22575$+$2046 & RBR 21                    & 51 Peg      & 113357 & 217014 & 2001.7340 & 2.87 & 154.5Z / 244.9N & 10.0$\pm$0.7\\

\enddata

\end{deluxetable}

\begin{deluxetable}{rllccccccc} 
\tabletypesize{\footnotesize}
\tablewidth{0pt}
\tablecaption{Orbital elements\label{orbital_elements}}
\tablehead{
\colhead{HD} &\colhead{Planet} &\colhead{Discoverer}  &\colhead{P}  &\colhead{a}       &\colhead{i}     &\colhead{$\Omega$} &\colhead{T$_0$} &\colhead{e} &\colhead{$\omega$}\\ 
\colhead{\#} &\colhead{}       &\colhead{Designation} &\colhead{(yr)} &\colhead{(\arcsec)} &\colhead{(\degr)} &\colhead{(\degr)}    &\colhead{(yr)}    &\colhead{} &\colhead{(\degr)} 
}
\startdata
19994 & \nodata & HJ 663 & 2029 &  9.87 &  104  & 97  & 2283  & 0.26  & 342\\
120136 & $\tau$ Boo & STT 270 & 996  &  8.01 & 49 & 174 & 2035 & 0.76 & 322\\
\enddata
\end{deluxetable}

\begin{deluxetable}{lll@{~}r@{~}lrrrrr} 
\tabletypesize{\scriptsize}
\tablewidth{0pt}
\tablecaption{Orbital Ephemerides\label{ephemeris}}
\tablehead{
\colhead{WDS} & 
\colhead{Planet} &
\multicolumn{3}{c}{Discoverer} &
\multicolumn{1}{c}{2015.0} &
\multicolumn{1}{c}{2020.0} &
\multicolumn{1}{c}{2025.0} &
\multicolumn{1}{c}{2030.0} &
\multicolumn{1}{c}{2035.0} \\
\colhead{Designation} & 
\colhead{~} & 
\multicolumn{3}{c}{Designation} &
\colhead{$\theta$~~~~~~~~$\rho$} & 
\colhead{$\theta$~~~~~~~~$\rho$} & 
\colhead{$\theta$~~~~~~~~$\rho$} & 
\colhead{$\theta$~~~~~~~~$\rho$} & 
\colhead{$\theta$~~~~~~~~$\rho$} \\
\colhead{~} & 
\colhead{~} & 
\multicolumn{3}{c}{~} &
\colhead{($\circ$)~~~~~~($''$)} & 
\colhead{($\circ$)~~~~~~($''$)} & 
\colhead{($\circ$)~~~~~~($''$)} & 
\colhead{($\circ$)~~~~~~($''$)} & 
\colhead{($\circ$)~~~~~~($''$)} \\
}
\startdata
03182$-$0112 & HD 19994 & HJ  &  663 &    & 196.4~~~2.178~ & 192.0~~~2.147~ & 187.4~~~2.129~ & 182.8~~~2.123~ & 178.2~~~2.131 \\
13473$+$1727 & $\tau$ Boo & STT &  270 &    &  62.6~~~1.774~ &  80.8~~~1.540~ & 103.3~~~1.446~ & 126.2~~~1.514~ & 145.5~~~1.697 \\
\enddata
\end{deluxetable}

\begin{deluxetable}{ccrrcc}
\tabletypesize{\footnotesize}
\tablewidth{0pt}
\tablecaption{Unresolved Stars\label{single_exostars}}
\tablehead{
\colhead{WDS} & \colhead{Planet} & \colhead{HD} & \colhead{HIP} &  \colhead{Epoch} & \colhead{FWHM} \\ 
\colhead{\#}  & \colhead{}       & \colhead{\#} & \colhead{\#}  &  \colhead{}      & \colhead{(\arcsec)} 
}
\startdata
\nodata &  \nodata      & 6434&      5054       & 2001.7453 & 0.14 \\

 \nodata        &  \nodata   & 8574 & 6643       & 2003.6891 & 0.08 \\
\nodata  & $\nu$ And          & 9826 & 7513      & 2001.7370 & 0.14 \\

\nodata  &             &        &     & 2001.7452 & 0.12 \\

\nodata   &109 Psc       & 10697  & 8159      & 2001.7452 & 0.11 \\

          & \nodata   & 12661 & 9683     & 2001.7453 & 0.13 \\

\nodata   &$\iota$ Her        & 17051 & 12653     & 2001.7454 & 0.12 \\

          &  \nodata  & 23596 & 17747     & 2003.7196 & 0.32 \\
03329-0927 &  $\epsilon$ Eri  & 22049& 16537   & 2003.7058 &  0.08\\
    &             &        &     & 2005.6471 &  0.12\\
    &             &        &     & 2005.6553 &  0.14\\
    &             &        &     & 2005.7837 &  0.14\\
    &             &        &     & 2005.7864 &  0.14\\
    \nodata      &  \nodata & 28185 & 20723      & 2003.7169 & 0.23 \\
\nodata           & \nodata  & 33636& 24205      & 2003.7360 & 0.12 \\
\nodata           & \nodata  & 38529 & 27253     & 2001.8987 & 0.17 \\

\nodata           & \nodata  & 40979 & 28767      & 2003.9740 & 0.18 \\
    &             &        &     & 2003.9796 & 0.12 \\
06332$+$0528 & \nodata  & 46375 & 31246     & 2004.0445 & 0.13 \\
 \nodata          &  \nodata & 49674 & 32916     & 2004.0860 & 0.27 \\
 \nodata          &  \nodata & 52265 & 33719      & 2004.0477 & 0.14 \\
 \nodata          &  \nodata & 68988 & 40687      & 2004.1299 & 0.16 \\
 \nodata          &  \nodata & 72659 & 42030     & 2004.1218 & 0.30 \\
 \nodata          &  \nodata & 74156 & 42723     & 2003.0074 & 0.08 \\
   &         &        &      & 2004.1216 & 0.33 \\
 \nodata          &  \nodata  & 75289 & 43177     & 2004.0313 & 0.39 \\
 \nodata & 55 Cnc         & 75732  & 43587    & 2002.0903 & 0.12 \\        
\nodata           &  \nodata  & 82943 & 47007    & 2001.9811 & 0.22 \\

  &            &        &     & 2002.1342 & 0.17 \\

\nodata            & \nodata  & 92788 & 52409    & 2002.0194 & 0.19 \\

\nodata  & 47 UMa           & 95128 & 53721   & 2001.9865 & 0.12 \\

\nodata           & \nodata  & 106252  & 59610   & 2003.2620 & 0.19 \\  
 \nodata            & \nodata   & 114729 & 64459   & 2003.2703 & 0.67 \\ 
\nodata            & \nodata   & 114783 & 64457   & 2003.2703 & 0.38 \\ 
\nodata            & \nodata  & 128311 & 71395    & 2003.3358 & 0.17 \\
\nodata  &23 Lib           & 134987 & 74500   & 2002.1069 & 0.17 \\

15249$+$5858 & $\iota$ Dra   & 137759 & 75458   & 2003.3221 & 0.08 \\
\nodata            & \nodata    & 141937 & 77740  & 2004.2560 & 0.12 \\

16010+3318  & \nodata & 143761 & 78459     & 2003.3169 & 0.24 \\ 

\nodata            & \nodata& 147513  & 80337 & 2003.5108 & 0.23\\
\nodata            & \nodata  & 150706 & 80902    & 2003.4945 & 0.38 \\
 & $\mu$ Ara            & 160691 & 86796   & 2002.3177 & 0.21 \\

\nodata           &  \nodata  & 168443 & 89844    & 2002.4954 & 0.22 \\

\nodata           &  \nodata   & 168746 & 90004   & 2002.5446 & 0.41 \\

\nodata           &  \nodata & 169830  & 90485    & 2002.5446 & 0.12 \\

   &            &        &     & 2002.6785 & 0.12 \\

\nodata           &  \nodata & 179949  & 94645    & 2002.5501 & 0.12 \\
\nodata           &  \nodata  & 190228 & 98714    & 2003.5249 & 0.12 \\
20036$+$2954   & GJ 777A  & 190360   & 98767    & 2002.6842 & 0.18 \\

\nodata           & \nodata  & 192263 & 99711    & 2002.5721 & 0.12 \\

\nodata           & \nodata   & 202206 & 104903  & 2002.5666 & 0.17 \\
\nodata           & \nodata   & 209458  & 108859  & 2002.5556 & 0.21 \\

  &           &        &     & 2002.6732 & 0.12 \\

\nodata           & \nodata   & 210277 & 109378  & 2002.5558 & 0.17 \\

   &           &        &     & 2002.6843 & 0.22 \\

22583$-$0224 &\nodata   & 217107 & 113421   & 2001.7370 & 0.19 \\
23393+7738   & $\gamma$ Cep & 222404 & 116727 & 2001.4963 & 0.10  \\

23419$-$0559 & \nodata & 222582 & 116906    & 2001.8653 & 0.39 \\
\enddata
\end{deluxetable}

\label{lastpage}


\begin{thebibliography}{}




\bibitem[Artymowicz \& Lubow(1994)]{artymowicz1994}
         Artymowicz, P. \&  Lubow, S.H. 1994, \apj, 421, 651 


\bibitem[Butler et al.(1997)]{butler1997}
         Butler, R.P., Marcy, G.W., Williams, E., Hauser, H., \& Shirts, P. 1997, \apjl, 474, L115

\bibitem[Chauvin et al.(2006)]{chauvin2006}
         Chauvin, G., Lagrange, A.-M., Udry, S., Fusco, T., Galland, F., Naef, D., Beuzit, J.L. \&  Mayor M. 2006, \aap, 456, 1165 


\bibitem[Cowley, Hiltner, \& Witt(1967)]{cowley1967}
         Cowley A.P., Hiltner W.A., \& Witt A.N. 1967, \aj, 72, 1334

\bibitem[Cox(2000)]{cox2000}
         Cox A.N. 2000, Allen's Astrophysical Quantities, Springer, New York, NY
 

\bibitem[Daemgen et al.(2009)]{daemgen2009}
         Daemgen, S., Hormuth, F., Brandner, W., Bergfors, C., Janson, M., Hippler, S., \& Henning, T. 2009, \aap, 498, 567

 
\bibitem[Desidera \& Barbieri(2007)]{desidera2007}
         Desidera, S., \& Barbieri, M. 2007, \aap, 462, 345

\bibitem[Dodson-Robinson et al.(2011)]{dodson-robinson2011}
         Dodson-Robinson, S.E., Beichman, C.A., Carpenter, J.M., \& Bryden, G. 2011, \aj, 141, 11
 
\bibitem[Eggenberger et al.(2004)]{eggenberger2004}
         Eggenberger, P. Udry, S., \& Mayor, M. 2004, \aap, 422, 247

\bibitem[Eggenberger et al.(2007)]{eggenberger2007}
         Eggenberger, A., Udry, S., Chauvin, G., Beuzit, J.-L. Lagrange, A.-M., Segransan, D., \&  Mayor, M. 2007, \aap, 474, 273 

\bibitem[Gray et al.(2001)]{gray2001}
         Gray R.O., Napier M.G., \& Winkler L.I. 2001, \aj, 121, 2148

\bibitem[Hale(1994)]{hale1994}
         Hale A.  1994, \aj, 107, 306
 
\bibitem[Hinkley et al.(2007)]{hinkley2007}
         Hinkley S. et al. 2007, \apj, 654, 633

\bibitem[Hinkley et al.(2011)]{hinkley2011}
         Hinkley S. et al. 2011, \apj, 726, 104

\bibitem[Leconte et al.(2010)]{leconte2010}
         Leconte J. et al. 2010, \apj, 716, 1551

\bibitem[Kley \& Nelson(2010)]{kley2010}
         Kley, W., \& Nelson, R.P. 2010, in Planets In Binary Star Systems (New York, Springer) 


\bibitem[van Leeuwen(2007)]{vanLeeuwen2007}
         van Leeuwen, F. 2007, \aap, 474, 653
 


\bibitem[Mason et al.(2011)]{mason2011}
         Mason, B.D., Hartkopf, W.I. Raghavan, D., Subsavage, J., Roberts, Jr., L.C., ten Brummelaar, T.A., \& Turner, N.H. 2011, \apj, Submitted

\bibitem[Mayor et al.(2004)]{mayor2004}
         Mayor, M., Udry, S., Naef, D., Pepe, F., Queloz, D., Santos, N.C., \& Burnet, M. 2004, \aap, 415, 391

\bibitem[Montes et al.(2001)]{montes2001}
         Montes D., Lopez-Santiago J., Galvez M.C., Fernandez-Figueroa M.J., De Castro E., \& Cornide M. 2001, \mnras, 328, 45

\bibitem[Mugrauer et al.(2004)]{mugrauer2004}
         Mugrauer M., Neuh\"{a}user R., Mazeh T., Guenther E., \& Fern\'{a}ndez M. 2004, Astron. Nachr. 325, 718

\bibitem[Mugrauer et al.(2005)]{mugrauer2005}
         Mugrauer M., Neuh\"auser R., Seifahrt A., Mazeh T., \&  Guenther E. 2005, \aap, 440, 1051
 
\bibitem[Mugrauer et al.(2007)]{mugrauer2007}
         Mugrauer, M., Neuh\"{a}user, R., \& Mazeh, T. 2007, \aap, 469, 755

\bibitem[Neuh\"{a}user et al.(2007)]{neuhauser2007}
         Neuh\"{a}user R., Mugrauer M., Fukagawa M., Torres G., \& Schmidt T. 2007, \aap, 462, 777

\bibitem[Patience et al.(2002)]{patience2002}
         Patience J. et al. 2002, \aj, 581, 654
  
\bibitem[Popovic \& Pavlovic(1996)]{popovic1996}
         Popovic G.M., \&  Pavlovic, R. 1996, Bull. Astron. Belgrade, 153, 57

\bibitem[Raghavan et al. (2006)]{raghavan2006}
         Raghavan D., Henry T.J., Mason B.D., Subasavage J.P., Jao W.-C., Beaulieu T.D., \&  Hambly N.C. 2006, \apj, 646, 523

\bibitem[Roberts \& Neyman(2002)]{roberts2002} 
         Roberts Jr., L.C.,  Neyman C.R. 2002, \pasp, 114, 1260

\bibitem[Roberts et al.(2005)]{roberts2005}
         Roberts Jr., L.C., Turner N.H., Bradford L.H, ten Brummelaar T.A., Oppenheimer B.R., Kuhn J.R., Whitman K., Perrin M.D.,  \&  Graham J.R. 2005, \aj, 130, 2262

\bibitem[Roberts(2011)]{roberts2011}
         Roberts Jr., L.C. 2011, \mnras, 413, 1200

\bibitem[Schnupp et al.(2010)]{schnupp2010} 
         Schnupp, C., et al. 2010, \aap, 516, A21 

\bibitem[Skrutskie et al.(2006)]{skrutskie2006}
         Skrutskie M.F. et al. 2006, \aj, 131, 1163

\bibitem[ten Brummelaar et al.(1996)]{tenBrummelaar1996}
         ten Brummelaar, T.A., Mason, B.D., Bagnuolo Jr., W.G., Hartkopf, W.I., McAlister, H.A., \&   Turner, N.H. 1996, \aj, 112, 1180

\bibitem[ten Brummelaar et al.(2000)]{tenBrummelaar2000}
         ten Brummelaar, T.A., Mason, B.D., McAlister, H.A., Roberts Jr., L.C., Turner, N.H., Hartkopf, W.I., \&   Bagnuolo Jr., W.G. 2000, \aj, 119, 2403

\bibitem[Torres(2007)]{torres2007}
         Torres, G. 2007, \apj, 654, 1095

\bibitem[Turner et al.(2008)]{turner2008}
         Turner, N.H., ten Brummelaar, T.A., Roberts Jr., L.C., Mason, B.D., Hartkopf, W.I., \&  Gies, D.R. 2008, \aj, 136, 554

\end{thebibliography}
\end{document}